\documentclass[11pt]{article}
\usepackage{amssymb}
\usepackage{amsmath}
\usepackage{simplewick}
\usepackage[all]{xy}
\usepackage[dvips]{graphicx}
\numberwithin{equation}{section}

\def\be{\begin{equation}}
\def\ee{\end{equation}}
\def\ba{\begin{align}}
\def\ea{\end{align}}
\def\beq{\begin{eqnarray}}
\def\eeq{\end{eqnarray}}

 \input{epsf}

 \usepackage{epsfig}

\begin{document}

\title{\Large{\bf The $AdS_3$ central charge in string theory}}
\author{Jan Troost$^{a}$ } \date{}
\maketitle
\begin{center}
   $^{a}$Laboratoire de Physique Th\'eorique \\
 Unit\'e Mixte du CNRS et
     de l'\'Ecole Normale Sup\'erieure \\ associ\'ee \`a l'Universit\'e Pierre et
     Marie Curie 6 \\ UMR
     8549 
\\ \'Ecole Normale Sup\'erieure \\
   $24$ Rue Lhomond Paris $75005$, France
\end{center}

 \begin{abstract}
   We evaluate the vacuum expectation value of the central charge
   operator in string theory in an $AdS_3$ vacuum. Our calculation
   provides a rare non-zero one-point function on a 
spherical worldsheet.  The evaluation involves the regularization both
   of a worldsheet ultraviolet divergence (associated to the infinite
   volume of the conformal Killing group), and a space-time infrared
   divergence (corresponding to the infinite volume of space-time).
   The two divergences conspire to give a finite result, which is the
   classical general relativity value for the central charge,
   corrected  in bosonic string theory 
by an infinite series of tree level higher derivative terms.
\end{abstract}

\newpage


\section{Introduction}
The boundary conditions on quantum gravity in three-dimensional
anti-de Sitter space can be chosen such that the asymptotic symmetry
group contains two copies of the centrally extended Virasoro
algebra. The value of the central charge was computed in
classical general relativity
\cite{Brown:1986nw}. The calculation can also be done using
holographic renormalisation \cite{Henningson:1998gx}. 

The space-time $AdS_3$ also arises as a factor in string theory
vacua. In string theory, the central charge in the space-time boundary
conformal field theory was computed in $AdS_3$ with
Neveu-Schwarz-Neveu-Schwarz flux in
\cite{Giveon:1998ns}\cite{Kutasov:1999xu}\cite{de
  Boer:1998pp}\cite{Giveon:2001up}\cite{Troost:2010zz} and for the
case of $AdS_3$ with Ramond-Ramond flux in \cite{Ashok:2009jw}.  In
string theory, the result is a central charge {\em operator} which can
take different values in different states of the theory
\cite{Giveon:2001up}. While the difference in central charge between
vacua separated by a long string has been computed
\cite{Giveon:1998ns}, the value in a given vacuum has not directly
been evaluated.  A natural assumption is that the
evaluation of the central charge operator in the $AdS_3$ vacuum should
give rise to the Brown-Henneaux central charge, at least at string
tree level and in the weak curvature limit.

The problem of the evaluation of the central charge in string theory
in $AdS_3$ with Neveu-Schwarz-Neveu-Schwarz flux is plagued both with
conceptual and technical difficulties. An attempt to repeat the
holographic Weyl anomaly calculation \cite{Henningson:1998gx} in
string theory would meet a basic obstacle.  The difficulty is that in
the derivation one uses the on-shell value of the bulk space-time
gravity action -- in string theory the on-shell action is zero
\cite{Tseytlin:1987ww}. 

Another difficulty becomes apparent when we realize that we are
supposed to evaluate a one-point function in string theory. In almost
all circumstances, a one-point function in string theory is zero
\cite{Moore:1985ix}, whether by symmetry, definition of regularization
scheme, or by the fact that one has an on-shell background. One formal
argument says that the one-point function will be divided by the
infinite volume of the worldsheet conformal Killing group (leaving one
point fixed), and will therefore vanish. Another argument says that
one-point functions must vanish due to conformal symmetry. These
statements are known to permit exceptions, for example for a
zero-momentum dilaton operator \cite{Liu:1987nz}, but not many. Every
exception is worth studying.

There is another fundamental reason to understand the $AdS_3$ central
charge in string theory. It is the central quantity to compute in a
check on the microscopic description of the degrees of freedom
responsible for the leading term in the macroscopic entropy of black
holes \cite{Strominger:1996sh}\cite{Strominger:1997eq}. 
As such, the value of the central
charge has propagated throughout the  literature.  We
believe it is important to compute it directly in string theory.

\section{The one-point function of the central charge operator}
Our goal is to calculate the one-point function of the central charge
operator in bosonic string theory on $AdS_3$ with
Neveu-Schwarz-Neveu-Schwarz flux. The calculation is universal for all
bosonic string theories including an $AdS_3$ space-time that
factorizes. 
Our computation is easily generalized to superstring theory
in which the supersymmetric
current algebra level is not renormalized.  We discuss the central
charge operator, first in the fully interacting picture, and then in
the asymptotically free field variables that will be handy in our
calculation.
\subsection{The central charge operator}
The classical string worldsheet action (with $\alpha'=2$) is given by:
\begin{eqnarray}
S_{int} &=& \frac{k}{2 \pi} \int d^2 z (\partial \tilde{\phi} \bar{\partial}
\tilde{\phi} + e^{2 \tilde{\phi}} \partial \bar{\gamma} \bar{\partial} \gamma),
\label{classaction}
\end{eqnarray}
where $\tilde{\phi}$ is the radial coordinate of the string, and the
coordinates $\gamma,\bar{\gamma}$ parameterize the planar sections of
the Poincar\'e patch of $AdS_3$ with Neveu-Schwarz-Neveu-Schwarz flux.
The fully interacting conformal field theory can be solved
\cite{Teschner:1997ft}. It contains primary operators of the form:
\begin{eqnarray}
\Phi_h &=& \frac{1}{\pi} ( |\gamma-x|^2 e^{\tilde{\phi}/{2}} + e^{- \tilde{\phi}/2})^{-2h},
\end{eqnarray}
and holomorphic and anti-holomorphic currents $J(x;z)$ and $\bar{J}
(\bar{x};\bar{z})$ which are linear combinations of the three
holomorphic and anti-holomorphic affine currents \cite{Kutasov:1999xu}.
There exists a scheme
\cite{Teschner:1999ug} in which the primary operators satisfy the
worldsheet operator
product expansion \cite{Kutasov:1999xu}:
\begin{eqnarray}
\Phi_1(x;z) \Phi_h(y;w) & = & \delta^2(x-y) \Phi_h(y;w)+ \dots
\label{threeOPE}
\end{eqnarray}
String theory on $AdS_3$ contains worldsheet vertex operators
corresponding to diffeomorphisms that fall of slowly at the boundary
of space-time.  These are graviton vertex
operators.  It can be argued that these operators are only formally
BRST exact \cite{Kutasov:1999xu}.  Using the worldsheet operator
product expansions of these vertex operators, one can prove that they
form two space-time Virasoro algebras in space-time, as they do in the
semi-classical limit \cite{Brown:1986nw}. A result of the calculation
is that the space-time Virasoro algebras contain a central charge
operator $C$ \cite{Kutasov:1999xu} which can indeed be 
proven to commute with the full algebra. The
central charge operator contains one right and one left oscillator
excitation, and a bulk-boundary propagator $\Phi_1$.
 In the scheme defined above, the
 central charge operator has the expression
 \cite{Kutasov:1999xu}\cite{Troost:2010zz}:
\begin{eqnarray}
C &=& -\frac{6}{k-2} \int d^2 z : \Phi_1 J \bar{J} : (z).
\label{C}
\end{eqnarray}
Although seemingly dependent on the boundary insertion point $(x,\bar{x})$,
it  can be shown that the derivative of the operator is BRST exact \cite{Kutasov:1999xu}. 
\subsection{Free field variables}
The vacuum expectation value of the central charge
operator is fixed by the asymptotic symmetry group,
and therefore it is sufficient for our purposes to compute near the
boundary of the $AdS_3$ space. In this region, there is a set of free
field variables that can be used to calculate quantities that depend on
a perturbative asymptotic interaction.
They are a real bosonic field
$\phi$ and a holomorphic $(\beta,\gamma)$ system of dimensions $(1,0)$
(as well as an anti-holomorphic $(\bar{\beta},\bar{\gamma})$ system).
The free field action is given by:
\begin{eqnarray}
S_0 & = & \frac{1}{4 \pi} \int d^2 z  \partial \phi \bar{\partial} \phi
\nonumber \\
& & 
+ \frac{1}{2 \pi} \int d^2 z (\beta \bar{\partial} \gamma
+ \bar{\beta} \partial \bar{\gamma}),
\end{eqnarray}
supplemented with a linear coupling of the scalar $\phi$ to the
worldsheet curvature such that the central charge of the theory is
equal to $c= 1 + \frac{6}{k-2} + 2 = \frac{3k}{k-2}$. The operator product expansions
of the free fields are:
\begin{eqnarray}
\phi (z) \phi(w) & \approx & - \log |z-w|^2
\nonumber \\
\beta(z) \gamma(w) & \approx & \frac{1}{z-w}.
\end{eqnarray}
We introduce the constant $Q=\sqrt{\frac{2}{k-2}}$.  The operator
$\beta \bar{\beta} e^{- Q \phi}$ is exactly marginal.  We can
therefore perturb the free conformal field theory with this
operator. We will study the perturbed theory with interaction term:
\begin{eqnarray}
S &=& S_0 - \frac{\mu}{2 \pi} \int d^2 z \beta \bar{\beta} e^{- Q \phi}.
\label{plusint}
\end{eqnarray}
Classically, the fields $\beta, \bar{\beta}$ can be integrated out to yield the
theory with action:
\begin{eqnarray}
S_{cl} &=& \frac{1}{4 \pi} \int d^2 z  \partial \phi \bar{\partial} \phi
\nonumber \\
& & 
+ \frac{1}{2 \pi} \int d^2 z \frac{1}{\mu} e^{Q \phi} \partial \bar{\gamma} \bar{\partial} \gamma
\end{eqnarray}
which we can rewrite as:
\begin{eqnarray}
S_{cl} &=& \frac{1}{4 \pi} \int d^2 z 
 \frac{4}{Q^2} \partial \tilde{\phi} \bar{\partial} \tilde{\phi}
\nonumber \\
& & 
+ \frac{1}{2 \pi} \int d^2 z \frac{1}{\mu} e^{2 \tilde{\phi}} \partial \bar{\gamma} \bar{\partial} \gamma.
\end{eqnarray}
At leading order in $1/k$, we can choose:
\begin{eqnarray}
\mu & \approx & 1/k,
\end{eqnarray}
and we recognize the string worldsheet action (\ref{classaction}) in an
$AdS_3$ background with purely Neveu-Schwarz-Neveu-Schwarz flux.

\subsection{A note on normalization}
The precise normalization of the central charge operator in equation
(\ref{C}) was computed in the fully interacting picture
\cite{Troost:2010zz}. We want to translate the operator, including its
overall normalization into the free field picture we will use in the
following. This translation was analyzed in detail in
\cite{Hosomichi:2000bm}\cite{Ishibashi:2000fn}.  We need to choose a
coefficient for the free field interaction term and normalize the
operators accordingly.

Suppose that we normalize
our operators in the free field approach naively as:
\begin{eqnarray}
\Phi_h^{free} & =& 
(|\gamma-x|^2 e^{Q \phi/2} + e^{-Q \phi/2})^{-2h},
\end{eqnarray}
and we pick the interaction term in equation (\ref{plusint}) as in
\cite{Hosomichi:2000bm}:
\begin{eqnarray}
\mu & \equiv & \frac{1}{k}.
\end{eqnarray}
Given these choices, the relation between the 
free field operators 
and the operators in the fully interacting formalism
is  \cite{Teschner:1999ug}\cite{Hosomichi:2000bm}\cite{Ishibashi:2000fn}:
\begin{eqnarray}
\Phi_h^{free} &=& E(h) \Phi_h
\end{eqnarray}
where
\begin{eqnarray}
E(h) &=& -b^4 (\frac{b^2}{\pi k})^{-h} \Delta(b^2)^{-2h+1} \Delta((2h-1)b^2).
\end{eqnarray}
We used the notations:
\begin{eqnarray}
\Delta(x) &=& \frac{\Gamma(x)}{\Gamma(1-x)} \nonumber \\
b^2 &=& \frac{1}{k-2}.
\end{eqnarray}
Thus, the interacting operator $\Phi_1$ that we wish to use
is given by the naive expression in the free field approach times
a factor of $E(1)^{-1} =  -(k-2)/(\pi k)$.
Thus, in the free field variable path integral, we must 
map the operator $\Phi_1$ to the expression:
\begin{eqnarray}
\Phi_1 &=&  - \frac{k-2}{\pi k} \frac{1}{
(|\gamma-x|^2 e^{Q \phi/2} + e^{-Q \phi/2})^2}.
\end{eqnarray}
\subsection{The central charge operator in the free field formalism}
We are finally ready to write down the normalized
central charge operator in the free
field formalism. We need the currents in the free field approach:
\begin{eqnarray}
J(x;z)
&=& - \beta (x-\gamma)^2 + 2(x- \gamma) \frac{1}{Q} \partial \phi
- k \partial \gamma,
\end{eqnarray}
and can then easily compute the asymptotic expression for the central
charge operator:
\begin{eqnarray}
C & = &  6k 
\int d^2 z  \delta^{(2)} (\gamma-x) \partial \gamma \bar{\partial} \bar{\gamma}
+ O(e^{- Q \phi}).
\end{eqnarray}
It will become manifest that 
only the leading term in the large radius limit
can contribute to the final result.
\subsection{The conformal field theory one-point function}
We first evaluate the central charge one-point function in the conformal
field theory, and then move on to embed the calculation in string theory.
When calculating the one-point function of
this operator using the free field operator products, we must 
 descend from the exponential of the action
 the appropriate number of interaction terms in order to find 
a non-zero result. In the case at hand, we find a non-zero result only when
we descend a single interaction term:
\begin{eqnarray}
\langle C \rangle_{CFT} & = & 6k \langle 
\int d^2 w  : \delta^{(2)} (\gamma-x) \partial \gamma \bar{\partial} \bar{\gamma} (w):
\frac{\mu}{2 \pi} \int d^2 z : \beta \bar{\beta} e^{- Q\phi} (z): \rangle
\nonumber \\
&  = & \frac{3  }{\pi}\int d^2 z \int d^2 w
\langle :\delta^{(2)} (\gamma-x)\partial \gamma \bar{\partial} \bar{\gamma} (w)
: :  \beta \bar{\beta} e^{- Q \phi} (z) : \rangle.
\end{eqnarray} 
We have the chiral correlator equality:
\begin{eqnarray}
\langle : \delta (\gamma-x)\partial \gamma (w):
:  \beta (z): \rangle
& = & \langle \partial_w  (\frac{1}{z-w} \delta (\gamma-x)) \rangle.
 \end{eqnarray}
Using this result, we can partially integrate and obtain:
\begin{eqnarray}
\langle C \rangle_{CFT} & = &
\frac{3  }{\pi}\int d^2 z \int d^2 w (2 \pi \delta^{(2)}(z-w))^2
 \delta^{(2)} (\gamma_0-x) e^{- Q \phi_0},
\end{eqnarray}
before integrating over the zero-modes.
The result is ultraviolet divergent on the worldsheet. 
We want to represent it in a form
that is more easily regularized. To that end, we rewrite:
\begin{eqnarray}
\int d^2 z \int d^2 w (2 \pi \delta^{(2)}(z-w))^2
&=&
\int d^2 z \int d^2 w \frac{1}{|z-w|^4}.
\end{eqnarray}
One way to read the integral is as the volume of the conformal Killing
group on the sphere that leaves one point fixed.
\subsection{The string theory one-point function}
Let's now move to the calculation of the string theory one-point function.
We need to take into account various modifications. Firstly, we need
to add the overall normalization of the path integral
\cite{Weinberg:1985tv}. We further need to divide out  the conformal field
theory correlator by the volume of the conformal Killing group of the sphere.
We must also still perform the integral over zero modes. And, we  need
to regularize the divergences. 

Firstly, we compute the ratio of the divergent conformal field theory
correlator and the volume of the conformal Killing group.
We must calculate the (inverse of the) following 
regularized ratio of integrals:
\begin{eqnarray}
& & \int d^2 z_1 d^2 z_2 d^2 z_3 \frac{1}{|z_1 - z_2|^2
|z_2 - z_3|^2|z_3 - z_1|^2 }
/
\int d^2 z d^2 w \frac{1}{|z - w|^4
} 
\nonumber \\
&  =& |z-w|^2 \int d^2 \xi \frac{1}{ |\xi -z |^2 |\xi-w|^2}  
\nonumber \\
& =&  4 \pi \log (1/\epsilon). 
\end{eqnarray} The points $z$ and $w$ are arbitrary points that fix
part of the $SL(2,\mathbb{C})$ symmetry of the formal triple integral.
In the last equality, we
used a short distance cut-off $\epsilon$ on the integral over the 
$\xi$-plane. The coefficient of the
logarithmic divergence is independent of the choice of points $z$ and
$w$. We note that the same regularization enters the calculation of the
 one loop mass renormalization
of string states in flat space \cite{Seiberg:1986ea}. It is also
reminiscent of the calculation of the space-time two-point functions in $AdS_3$ 
\cite{Maldacena:2001km}.

Next, we note that the overall normalization of the bosonic string
path integral including ghosts
 is such that it provides us with an extra factor:
\begin{eqnarray}
\frac{8 \pi}{\alpha' g_s^2}.
\end{eqnarray}
This constant is fixed by unitarity 
\cite{Weinberg:1985tv}\cite{Polchinski:1998rq}.
Putting all these results together,
the full expression for the one-point function including the zero mode
integral (and at $\alpha'=2$) becomes:
\begin{eqnarray}
\langle C \rangle_{string} &= & \frac{3 V^{\perp} }{ g_s^2 
\pi \log (1/\epsilon)}
 \int_{}^{\phi_0^{IR}}
d \phi_0 \int d \gamma_0 \int d \bar{\gamma}_0 
e^{Q \phi_0}
 \delta^2(x-\gamma_0)
e^{- Q \phi_0}.
\end{eqnarray}
The transverse volume factor $V^{\perp}$ arises from the
integration over the zero modes orthogonal to $AdS_3$.  Note that we
also took into account the linear dilaton contribution to the $\phi_0$
zero-mode integral on the sphere, and that we introduced a bulk
infrared cut-off $\phi_0^{IR}$ on the radial integration region.
We take the lower limit on the integration region to be close enough to 
the boundary for our approximation to the central charge operator to be valid
(e.g. $\phi_0 >> 1/Q$).
This is sufficient to isolate the bulk infrared divergent contribution
of interest.

In bosonic string theory, we have the following relation between
the string coupling and the dimensionally reduced Newton coupling constant
\cite{Polchinski:1998rq}:
\begin{eqnarray}
\frac{1}{g_s^2} 
&=& \frac{\pi}{2 G_N^{3} V^{\perp}}.
\end{eqnarray}
Using this relation, as well as $Q = \sqrt{2/(k-2)}$ and $\alpha'=2$,
we finally find:
\begin{eqnarray}
\langle C \rangle_{string} & = &
\frac{3 \sqrt{(k-2)\alpha'} }{2 G_N^3}
 \frac{ \log e^{\frac{Q}{2} \phi_0^{IR}}}{  \log 1/\epsilon}.
\end{eqnarray}
We wrote the argument of the logarithm in the numerator as a function
of the space-time infrared length scale (as is clear
from the expression for the space-time metric). We can now justify the 
regularized expression:
\begin{eqnarray}
\langle C \rangle_{string} & = &
\frac{3 \sqrt{(k-2)\alpha'} }{2 G_N^3}
\end{eqnarray}
in various ways. First of all, the logarithmic divergences we
identified are universal, and so is their coefficient. Moreover, a
definition of closed string one-point functions can be given in terms
of the logarithmic derivative with respect to a worldsheet ultraviolet
cut-off \cite{Tseytlin:1988tv}.  It is also natural to identify the
logarithmic dependence on the space-time infrared cut-off as the
conformal anomaly. More powerfully, by holography, we can kill two
birds with one stone. A holographic interpretation for our
regularized expression is to view the logarithmic divergences as
cancelling one another directly. Indeed, the worldsheet ultraviolet
cut-off is akin to a boundary ultraviolet cut-off, which in turn is
holographically dual to a bulk infrared cut-off
\cite{Susskind:1998dq}. As a final argument for our final
equation we note that our
identification reproduces the small curvature general relativity limit
\cite{Brown:1986nw}. Our result moreover contains an
infinite set of tree level higher derivative corrections which
are present
in bosonic
string theory.
\section*{Acknowledgements}
Our work was supported in part by the grant ANR-09-BLAN-0157-02. We
would like to thank the referee for suggestions to improve our
paper. It is a pleasure to thank Costas Bachas for an interesting
discussion.

\end{document}